\documentclass[preprintnumbers,twocolumn,amsmath,amssymb]{revtex4}
\usepackage{graphicx}
\usepackage{dcolumn}
\usepackage{bm}
\usepackage{multirow}
\usepackage[utf8]{inputenc}
\usepackage[english,russian]{babel}

\begin{document}

\title{Coulomb corrections and thermo-conductivity of a dense plasma}

\author{S. I. Glazyrin}
\altaffiliation[Also at ]{Moscow Institute for Physics and Technology, Dolgoprydnyi, Russia}
\email{glazyrin@itep.ru}

\author{S. I. Blinnikov}
\email{Sergei.Blinnikov@itep.ru}
\affiliation{Institute for Theoretical and Experimental Physics, Moscow, Russia}

\begin{abstract}
  We point out that confusion sometimes arises when using a chemical potential in plasma
  with Coulomb interactions. The results of our consideration are applied to
  the discussion of nuclear reactions screening. Finally, we present a transparent
  derivation of the thermal conductivity coefficient of a degenerate electron gas.
\end{abstract}


\maketitle





\section{Introduction}

This paper considers some questions that are important when nuclear
flames in a dense plasma are considered. These questions can be applied
to many systems, for example, to supernovae of type Ia.
If one considers papers dealing with  nuclear statistical equilibrium
(NSE) or with nuclear reactions screening, one can
note that authors use different definitions of chemical potential and different expressions for
Coulomb corrections. Some of these definitions are general and some
are accurate only in particular cases. Inaccurate definitions
can sometimes lead to the wrong conclusions and misleading understanding of a physical situation.
The aim of this paper is to arrange all
definitions and basic equations into a physical perspective and to discuss some paradoxes arising
from inaccurate definitions.
The second part of the article deals with kinetic properties of matter in
white dwarfs. The question was considered in the 1960--1970s, \cite{FlowersItoh1976,YakovlevUrpin1980} and
references therein. One of the most important parameter for supernovae Ia is
thermo-conductivity of degenerate electron gas.
A method for calculating thermal conductivity
coefficient was described in \cite{FlowersItoh1976} but final expression was not obtained;
whereas in \cite{YakovlevUrpin1980} a correct final expression was
obtained but derivation is not so rigorous.
There are some improvements in \cite{Yak1987,CassisiEtAl2007,Ichimaru1993}.
Here we present a straightforward and more transparent derivation,
based on ideas from \cite{FlowersItoh1976}.

\section{Chemical potential}

A general definition of a chemical potential $\mu$ is contained in the differential
of internal energy $E$,
\begin{equation}
  dE=TdS-PdV+\mu dN .
\end{equation}
We use standard notations here and $\mu dN$ may be a sum for several components:
$\mu dN \rightarrow \sum\limits_i \mu_i dN_i$.
The differential of the Helmholtz free energy $F\equiv E-TS$ is written as follows
\begin{equation}
  dF=-SdT-PdV+\mu dN  ,
\end{equation}
and thus it means that
\begin{equation}
  \mu=\left(\frac{\partial F}{\partial N}\right)_{T,V}  .
\label{eq:muF}
\end{equation}

Let us consider the Coulomb  free energy of a set of ${N_i}$ nuclides:
\begin{equation}
  F_{\rm nuc}=\sum\limits_i N_i \Phi_i(T,{n_k}) .
\end{equation}
Here $\Phi_i$ is the free energy per nucleus $i$ and it may depend
not only on the concentration,
\begin{equation}
n_i=\frac{N_i}{V},
\label{eq:concen}
\end{equation}
of the $i$-th nucleus but also on concentrations of some other nuclei
$n_k$. We have by definition
\begin{eqnarray}
\mu_i={\partial F_{\rm nuc}\over \partial N_i}\bigg |_{T,V}
=\Phi_i +\sum_k n_k{\partial \Phi_k\over \partial n_i}\bigg |_T\;.
\label{eq:chempo}
\end{eqnarray}
Then we find, for example, the pressure of nuclide gas:
\begin{eqnarray}
P_{\rm nuc}=-{\partial F_{\rm nuc}\over \partial V}\bigg |_{T,\{N_k\}}=
\sum_k\mu_kn_k -F_{\rm nuc}/V,
\label{eq:press}
\end{eqnarray}
which is consistent with the definition of the grand thermodynamic
potential, $\Omega = - PV$.
%

A convenient parameter of plasma non-ideality is
\begin{eqnarray}
\Gamma_i  \equiv {1\over k_{\rm B}T}\biggl(
{e^2Z_i^2\over a_{i}}\biggr) ={e^2Z_i^2\over k_{\rm B}T} \biggl(
{4\pi \over 3} n_i\biggr)^{1/3} .
\label{eq:Gamma}
\end{eqnarray}
According to accurate Monte-Carlo (MC) simulations,
a very good  theoretical model for Coulomb corrections in a neutral one-component plasma
 is Wigner-Seitz approximation (in the first order of $\Gamma$),
see references, e.g. in \cite{Kraeft1986}.
In this approximation, the correction to the free energy of a system is
\begin{equation}
  \Delta F^C = -0.9 N_i \Gamma_i k_B T,
\label{eq:DeltaFC}
\end{equation}
where $a_i$ is an average distance between ions of type $i$,
\begin{equation}
  a_i=\left(\frac{3}{4\pi}\frac{V}{N_i}\right)^{1/3} .
\end{equation}
Such a definition of $a_i$ is often used (see e.g. \cite{PDB06}), but
we point out in advance of full discussion that this is correct only for one-component plasma.

A good correspondence of this simple model with MC simulation one may find, e.g. in
the results by \cite{DewittSlattery1999},
where the main correction to
free energy is approximated by $f=a\Gamma+...$ and $a=-0.899172$.

So in the case of Wigner--Seitz approxomation the Coulomb correction
to chemical potential from equations (\ref{eq:muF})
and (\ref{eq:DeltaFC}) is
\begin{equation}
  \Delta\mu^C=-1.2 \frac{Z_i^2e^2}{a_i}=-1.2\Gamma_i k_B T,
  \label{eq:muCoulCorr}
\end{equation}
the result which has coefficient that is 4/3 times larger than the
Wigner-Seitz's 0.9 because  $a_i$ depends on $N_i$ too.
Here and below we will focus our attention on the
numerical coefficient in the correction; thus the coefficient is very important.

Sometimes $a_i$ is introduced in another way, through electron density
\begin{equation}
  a_i=\left(\frac{3}{4\pi}\frac{Z_i}{n_e}\right)^{1/3} ,
\end{equation}
and the resulting expression for the Coulomb corrections is
\begin{equation}
  \Delta F^C = -0.9 N_i Z_i^{5/3}e^2\left(\frac{4\pi}{3}n_e\right)^{1/3} \; .
  \label{eq:FcByNe}
\end{equation}

At the first glance such an expression leads to the coefficient 0.9 in~(\ref{eq:muCoulCorr}),
but in this case $n_e$ depends on the number of ions. And if we take
out one ion from our system then, for preserving electro-neutrality,
we should take out $Z$ electrons; hence, $n_e$ changes so that we obtain the correct answer~(\ref{eq:muCoulCorr}).

These considerations have direct implications for NSE
which is essentially reduced to a Saha-like equations for abundances of nuclides.
A correct form of these equations should be derived from correct expressions
for chemical potentials in equilibrium.
However, it is also used sometimes in another approach, when Coulomb corrections
(with the coefficient 0.9)
are introduced directly to the binding energies of reacting nuclei.
This approach also seems natural, at least while nuclides form an almost ideal
gas (see, e.g., a widely cited paper  \cite{maz79}), but the results
for nuclear abundances will differ from those obtained with
the correction (\ref{eq:muCoulCorr}) for chemical
potential.

We can derive the result (\ref{eq:muCoulCorr}) from a more general expression.
If $Q_i$ denotes the Coulomb correction to the free energy
and has the functional form
\begin{eqnarray}
 Q_i \equiv k_BTf_0(\Gamma_i),
\label{eq:mazurek}
\end{eqnarray}
then we find the Coulomb contribution to the chemical potential as
\begin{eqnarray}
\mu_i^{\rm Coul}=k_BT\biggl[f_0(\Gamma_i)+{1\over 3}{\partial f_0 \over
\partial \ln \Gamma_i}\biggr] ,
\label{eq:Coulmu}
\end{eqnarray}
this agrees with (\ref{eq:muCoulCorr}) in case of the Wigner-Seitz approximation.
But we want to point out that it contradicts equation (17) of \cite{maz79} often used
in NSE studies.
Our expression (\ref{eq:Coulmu}) does not coincide
also with equation (97) of the famous review by \cite{yak89}.
It was fount in \cite{yak89} that $\mu_i=k_BT f_0(\Gamma_i)$,
but this expression leads to zero correction to $P_{\rm nuc}$ in equation (\ref{eq:press}).
On the other hand, our equation (\ref{eq:Coulmu}) gives the expected correction to
$P_{\rm nuc}$ after substitution into equation (\ref{eq:press}).

It is a very important point in our reasoning, so let us consider it
carefully. As will be shown in section~\ref{sec:screening}, a more appropriate expression
for Coulomb corrections is (\ref{eq:FcByNe}), and not~(\ref{eq:Gamma})-(\ref{eq:DeltaFC}).
But if we use the assumption of electro-neutrality, the numbers of
electrons and ions are directly related. It is correct
in case when we consider our system on a timescale that is much larger than
plasma frequency (like in NSE) -- so we cannot separate electrons and
ions. In this case, the chemical potential is defined as
\begin{equation}
  \mu=\left.\frac{\partial F}{\partial N_i}\right|_{T,V,N_{\rm
      ions},{\rm electro-neutr.}},
  \label{eq:correct_mu}
\end{equation}
and for one-component mixture, the answer is~(\ref{eq:muCoulCorr}).
Here the full correction could be ascribed to ions if electrons are
considered as background or to electrons if ions are background.
But when we consider kinetic properties of our medium, we study it on
timescales when electro-neutrality could be violated --- so we should
consider electrons and ions separately, like the corrections
\begin{eqnarray}
 \Delta\mu_i^C&=&\left.\frac{\partial \Delta F_C}{\partial N_i}\right|_{T,V,N_e}=-0.9
  Z_i^{5/3}e^2\left(\frac{4\pi}{3}\frac{N_e}{V}\right)^{1/3}\; ,\nonumber\\
  \Delta\mu_e^C&=&\left.\frac{\partial \Delta F_C}{\partial N_e}\right|_{T,V,N_i}=-0.3 \frac{N_i}{N_e}
  Z_i^{5/3}e^2\left(\frac{4\pi}{3}\frac{N_e}{V}\right)^{1/3}
  \label{sys:muimue}
\end{eqnarray}
these corrections should be applied to~(\ref{sys:i_q}) below (there, in a kinetic
case, all thermodynamic quantities have an approximate meaning, and
have to be carefully defined similar to that in \cite{ll10}).
Here we should emphasize one important point. Coulomb corrections
in dense plasma arise due to electron-ion interactions because all
ions are screened by electron clouds.
The notation $F_{ii}$ that is used in many works is a bit misleading
because it is not pure ion-ion interaction.

In the book by \cite{HPYa07} the authors rely mostly on free energy
$F$ instead of $\mu_i$. But in many cases correct knowledge of $\mu$ is required.
Actually, the expression which is equivalent to (\ref{eq:Coulmu}) appeared already in
\cite{BST1966}. In the paper \cite{DewittGraboskeCooper1973}, the
result for the $\mu$ coefficient is close to 1.2, but it was obtained
using Monte-Carlo sumilation and rough expansion.
However,
the argument on the role  of the chemical potentials and debates on approaches to them
still persists. For example,  in \cite{Brown06} the difference of their use of
chemical potentials have been discussed
with that of \cite{DewittGraboskeCooper1973}. Essentially, in
\cite{Brown06} it was found that the
Coulomb correction found in \cite{DewittGraboskeCooper1973} for a classical plasma is
applicable to quantum plasma as well.
(See also \cite{Brydges99} for a detailed review on modern perspective on Coulomb
systems.)

Since there is some controversy on the form of Coulomb
corrections in the literature, in the paper \cite{BPRS2009} the
results for the Coulomb correction from equation (\ref{eq:mazurek}), as was
taken by \cite{maz79}, with the effect of the additional  term in
equation (\ref{eq:Coulmu}) have been compared.
The effect of the new term in (\ref{eq:Coulmu}) on the proton fraction (which is very important in collapse
computations) is visible only at highest density
but still very small.
A bit more pronounced effect is obvious 
for the average mass of heavy nuclei again at highest density when the NSE model
is near the border of applicability.
However, a more detailed study is
needed here along the lines undertaken recently by \cite{nad05}.

\section{Screening of nuclear reactions}
\label{sec:screening}

A very important application of the previous reasoning is in the field of
a classical nuclear burning screening.
Consider the regime of strong screening $\Gamma\ge 1$. According to
\cite{DewittGraboskeCooper1973} the expression for screening factor in this case can be
written as
\begin{equation}
  \exp\left(H_{12}(0)\right)=\exp(-\left\{
  \mu(Z_1+Z_2)-\mu(Z_1)-\mu(Z_2)\right\}/T) .
  \label{eq:Hmu}
\end{equation}
Here and below $k_B \equiv 1 $.
So theoretically, using the above definition of a chemical potential, the screening factor is
\begin{equation}
  H_{12}(0)=1.2\left(\Gamma_{Z_1+Z_2}-\Gamma_{Z_1}-\Gamma_{Z_2}\right) .
  \label{eq:Hgamma_mu}
\end{equation}
In order not to create confusion, let us consider a commonly examined situation when
$Z_1=Z_2=Z$. We obtain in the leading order
(here we will discuss only the leading order of corrections)
\begin{equation}
  H_{12}(0)=1.410\Gamma_{Z}
  \label{eq:HGammaEqZ} ,
\end{equation}
(here $1.410=1.2(2^{5/3}-2)$).
On the other hand if we consider the definition of screening factor using
free energy,
\begin{equation}
  H_{12}(0)=-\left(F(2Z)-2F(Z)\right)/T,
  \label{eq:HF}
\end{equation}
and using~(\ref{eq:DeltaFC}) obtain
\begin{equation}
  H_{12}(0)=0.9\left(\Gamma_{2Z}-2\Gamma_{Z}\right)=
  1.057\Gamma_Z.
  \label{eq:Hgamma_F}
\end{equation}
In the case of $\Gamma\gg 1$ the difference in~(\ref{eq:Hgamma_mu})
and~(\ref{eq:Hgamma_F}) can be extremely high. In some works the chemical
potential has coefficient 0.9 and so using two definitions of $H_{12}$
the answer~(\ref{eq:Hgamma_F}) is obtained.

First of all, let us consider the physical sense of a chemical potential. Equation
(\ref{eq:Hmu}) tells us that we take out from our system two ions with a charge $Z$ and
put back one ion with a charge $2Z$. From the chemical point of view, ions with
$Z$ and $2Z$ are different elements and the average distance between $Z$-ions
changes. So if we recall that coefficient 1.2 arises from the  change of  $a_i$,
then~(\ref{eq:Hgamma_F}) is incorrect.
But for finding a proper screening correction one should stand on  electromagnetic force 
point of view: these are elements of the same nature -- and there should be only one distance between
reactants.

At this point a reader can object in the following way: a nuclear reaction
is a fast process, but the effect
we are discussing now exists because of a global (on scales of our system) variation
of the average distance between ions: this is a slow process. Can it really change the
repulsion potential --- a barrier, our ion is penetrating through? Theory
of such corrections rely on a statistical averaging
\cite{DewittGraboskeCooper1973,Jancovici1977} and operate
with average quantities, so in such a theory we should consider these effects.

On the other hand, if we consider definition of free energy using
$n_e$~(\ref{eq:FcByNe}), in nuclear reactions, the electron number
density $n_e$ stays constant, unlike the situation when we define chemical
potential and take one ion with its electron cloud from our system.
The average distance $a_i$  does not change in these terms, and then the coefficient
0.9 should stay in~(\ref{eq:Hgamma_mu}).
We see that there is some paradox in this situation: one reasoning
leads to 0.9, another to 1.2.

Let us resolve the paradox.
Expression (\ref{eq:Hgamma_F})
used by scientific community is correct. The reason of the misunderstanding is
 the definition of $a_i$ which is sometimes incorrect.
What is the Wigner-Seitz model? This model tells us that the most significant Coulomb
corrections are due to the interaction of ions with electrons (and defining
these correction through the average distance between ions may result in a confusion of
the reader). Electron gas is
uniformly spread over the whole volume (we omit such effects as electron polarization
etc) and we should define $a_i$ for an ion $Z_i$ as a radius around the ion, which contains
electron charge $(-Z_i)$:
\begin{equation}
  a_i\equiv \left(\frac{3}{4\pi}\frac{Z_i}{n_e}\right)^{1/3} .
\end{equation}
Here, in general, $a_i$ is not the average distance between ions:
this would be true only if we had a one-component mixture ---
what is undoubtedly wrong in nuclear reactions, our mixture after reaction is multicomponent.
So the correct definition of $a_i$ is a radius of electron cloud with charge $(-Z_i)$.
Such an electron coat screens nuclear repulsion potential
and one ion from another ---
there is no direct Coulomb interaction between ions. As we have mentioned above, $n_e$ is
constant, so the correct leading order in screening is
\begin{equation}
  H_{12}(0)=1.057\Gamma_{Z} .
\end{equation}
Finally we could write the full classical correction from MC simulation with
equal charges
(see e.g. \cite{DewittSlattery1999})
\begin{eqnarray}
  H_{12}(0)&=1.056299\Gamma_Z+1.039957\Gamma_Z^{0.323064}-\nonumber\\
  &-0.545823\ln\Gamma_Z
  -1.036319 .
\end{eqnarray}
Let us emphasize one moment. When defining chemical potential, we
should obey the electro-neutrality law and take one ion from the system
with its electron cloud. And in nuclear reactions nothing is taken
from the system, so the chemical potential has
another physical meaning, and the correct definition is done by free
energy~(\ref{eq:HF}).
A very bright demonstration of this point is the famous
review by Ichimaru~\cite{Ichimaru1993} with its section C on
``Chemical potentials'': only free energies are discussed, and $\mu$
symbol does not appear there.

\section{Thermal Conductivity}

It is commonly supposed that write dwatfs (WD) are progenitors for
supernovae Ia. On the initial stage of explosion, thermo-conductivity plays
a significant role as it drives the flame front of thermonuclear burning
\cite{HillebrandtNiemeyer2000}.
Typical parameters of matter in presupernovae WD are $T\sim 10^9$~K
and $\rho \sim 10^9$~g/cm$^3$.
Chemical composition is a mixture $^{12}{\rm C}$+$^{16}{\rm O}$.
For these conditions the electron gas is relativistic and degenerate
$\epsilon_F\gg k_BT$, $p_F\gg m_ec$, and the parameter of non-ideality (\ref{eq:Gamma}) $\Gamma\sim 1$.

Let us consider an electron gas which is described by its distribution function in
a phase space $f({\bf p},{\bf x},t)$. Boltzmann equation for evolution of
such a function is the following (in this section we will use units
in which $c=\hbar=k_B=1$ and in the end we will restore dimensionality of variables):
\begin{equation}
\frac{df}{dt}={\rm St} f = -\int w ({\bf p},{\bf p'})\left[
f(1-f')-f'(1-f)\right]\frac{2d^3
p'}{(2\pi)^3}.
\label{eq:Bolzman}
\end{equation}
Here we assume that the scattering of electrons occurs only on ions, and the distribution function of the latter does not change. Under our conditions the ions are
non-degenerate, non-relativistic and
have large mass in comparison with electrons mass; they can be treated as being at rest.
An approach which considers the scattering on one ion (separately from other ions)
is incorrect when $\Gamma\gg 1$ because of the correlation in ion positions.
Thus we shall consider a situation when $\Gamma\le 1$ and suppose that our answer has a right limit $\Gamma \rightarrow 1$ (finally we want the answer for $\Gamma\sim 1$).
This is true because when $\Gamma\ll 1$ we have an ideal gas behavior
of ions. The case $\Gamma\sim 1$ is called Coulomb liquid: there is no significant correlation
in positions and full correlation sets in the solid state when $\Gamma\sim 170$ (see~\cite{vHorn1969,SlDooDew1980,JP1996}).

Cross-section of scattering on an ion at rest (with potential $A_0(r)= Z e/r$) is written
as follows~\cite{ll4}
(we will use a short definition $d{\bf p'}\equiv {2d^3 p'}/{(2\pi)^3}$):
\begin{equation}
\frac{d\sigma}{d{\bf p'}}=\pi e^2\delta(\epsilon_{p'}-\epsilon_p)|A_0({\bf
p'}-{\bf p})|^2\left(1-\frac{{\bf
q}^2}{4\epsilon_p^2}\right)\frac{\epsilon_p}{|{\bf p}|},
\label{eq:sigma}
\end{equation}
where $A_0({\bf q})=4\pi Ze/{\bf q}^2$.
The $\delta$-function tells us that the scattering
is elastic.
The comparison  of e-i and e-e cross-sections gives
$\sigma_{ei}/\sigma_{ee}\sim Z^2\gg 1$, see e.g. \cite{ll4}, so we omit the e-e scattering here.

As usual, we consider a small deviation from the  equilibrium distribution function:
\begin{equation}
f_p=f^0_p-\frac{\partial f^0_p}{\partial \epsilon_p}\phi_p=
f^0_p+\frac{1}{T}f^0_p(1-f^0_p)\phi_p,
~~~~f^0_p=\frac{1}{e^{(\epsilon_p-\mu)/T}+1}.
\label{eq:Distrib0}
\end{equation}

So Boltzmann equation in the case of thermo-conductivity looks like
\begin{equation}
-\frac{\partial f^0}{\partial\epsilon}\frac{\epsilon-w}{T}({\bf v}\nabla)T=
-\frac{1}{T}\int w({\bf p},{\bf p'})(\phi-\phi')f^0 (1-f^0)\frac{2d^3p'}{(2\pi)^3}.
\end{equation}
In the case of degenerate electron gas, $sT\ll \mu$, so we could substitute
$\mu$ instead of $w$.
Let us define the abbreviation $\Gamma_{p\rightarrow p'}\equiv w({\bf
p},{\bf p'})f^0(1-f^0)$. After multiplication on $\phi$ and integration
over the whole $p$, we can reduce the Boltzmann equation to the form
\begin{equation}
\frac{j_E^2}{\kappa T}=\frac{1}{2T}\int(\phi-\phi')^2\Gamma_{p\rightarrow
p'}\frac{2d^3 p}{(2\pi)^3}
\frac{2d^3 p'}{(2\pi)^3},
\label{eq:EntropyMix}
\end{equation}
where energy flux is defined as
\begin{equation}
{\bf j_E}=\int(f-f^0){\bf v}(\epsilon-\mu)\frac{2d^3 p}{(2\pi)^3} ,
\end{equation}
and the thermo-conductivity $\kappa$ is defined in the following way:
\begin{equation}
 {\bf j_E}=-\kappa\nabla T .
\end{equation}
Equation (\ref{eq:EntropyMix}) has exact physical meaning -- it can be shown (see~\cite{Ziman1960})
that the rhs of (\ref{eq:EntropyMix}) is a rate of entropy generation
by particles' scattering. On the lhs is
the entropy generation rate through conductivity.
The following method of coefficient determination is good -- we could
substitute arbitrary functions $\phi$ and get answers for
$\kappa$. There exists a theorem that the maximum value of $\kappa$ for
different $\phi$ is mathematically exact.

This is a simple case, namely, isotropic. There is a limited set of quantities in
our task; that is why we could guess the answer
\begin{equation}
\phi=A(\epsilon_p-\mu) p_\alpha\partial_\alpha T .
\end{equation}

The previous expression can be simply explained: it should be proportional
to the temperature gradient, the only physical quantity that can create a scalar with
combination with gradient
is a momentum. The factor $(\epsilon_p-\mu)$ was introduced because all macroscopic parameters
should be defined through $f_0$ but not through the correction proportional to $\phi$. 
The factor $A$ is a dimensional constant, but it will be cancelled out immediately 
after the insertion of $\phi$ into (\ref{eq:EntropyMix}), so we omit it now.

Let us consider the integration
\begin{equation}
{\bf j_E}=\frac{1}{T}\int (\epsilon_p-\mu){\bf v} f_p^0(1-f_p^0)\phi d{\bf p}.
\end{equation}

Choosing for definiteness  $\partial_\alpha T=\left(0,0,\nabla T\right)$ and
using the fact of high degeneracy of electrons
(therefore $\int\limits_0^\infty
\psi(\epsilon)\frac{\partial f_0}{\partial\epsilon}d\epsilon=
-\psi(\mu)
-\frac{\pi^2}{6}T^2\psi''(\mu)-\frac{7\pi^4}{360}T^4
\psi''''(\mu)+...$) we obtain
\begin{equation}
j_{E_z}=\frac{1}{9}(\nabla T)T^2(\mu^2-m_e^2)^{3/2} .
\label{eq:j_E_z}
\end{equation}
The rhs of equation~(\ref{eq:EntropyMix}):
\begin{equation}
\mbox{RHS}=\int(\phi-\phi')^2\Gamma_{p\rightarrow p'}\frac{2d^3 p}{(2\pi)^3}
\frac{2d^3 p'}{(2\pi)^3},
\end{equation}
after all substitutions
\begin{widetext}
\begin{equation}
\mbox{RHS}=\int(\epsilon_p-\mu)^2 q_\alpha (\partial_\alpha T)
q_\beta(\partial_\beta T)\pi e^2\delta(\epsilon-\epsilon')|A_0|^2
\left(1-\frac{{\bf q}^2}{4\epsilon_p^2}\right)\frac{\epsilon_p}{p}\cdot
\cdot f_p^0(1-f_p^0)n_i v\frac{2d^3 p}{(2\pi)^3}\frac{2d^3 p'}{(2\pi)^3} .
\end{equation}
Doing the same integration,
\begin{equation}
\mbox{RHS}=\frac{1}{18\pi}e^2 n_i T^3 \mu(\mu^2-m_e^2)^{1/2}
(\nabla T)^2\frac{\epsilon}{p}
\int\limits_0^{2p} q^3 |A_0(q)|^2
\left(1-\frac{{\bf q}^2}{4\mu^2}\right)dq.
\end{equation}
\end{widetext}

Let us denote Coulomb logarithm (recalling that $A_0(q)=4\pi Z e/q^2=4\pi Z e \varphi(q)$)
\begin{equation}
\Lambda_{ei}=\int\limits_0^{2p} q^3 |\varphi(q)|^2
\left(1-\frac{{\bf q}^2}{4\mu^2}\right)dq ,
\label{eq:CoulLog}
\end{equation}
\begin{equation}
\mbox{RHS}=\frac{8\pi}{9}Z^2 e^4 n_i T^3 \mu(\mu^2-m_e^2)^{1/2}
(\nabla T)^2\frac{\epsilon}{p}\Lambda_{ei} .
\label{eq:RHSfinal}
\end{equation}
After the substitution of~(\ref{eq:j_E_z}) and~(\ref{eq:RHSfinal}) in~(\ref{eq:EntropyMix}),
and defining $v_F \equiv p/\epsilon $, we obtain
\begin{equation}
\kappa=\frac{1}{36\pi}\frac{T(\mu^2-m_e^2)^{5/2}v_F}{Z^2 e^4 n_i\mu\Lambda_{ei}},
\label{eq:Kappa}
\end{equation}
recovering dimension
\begin{equation}
\kappa=\frac{1}{36\pi}\frac{k_B^2 T(\mu^2-m_e^2 c^4)^{5/2}v_F}{(\hbar c)^3 Z^2 e^4 n_i\mu\Lambda_{ei}}.
\label{eq:KappaRasm}
\end{equation}
This is the same result as calculated in~\cite{YakovlevUrpin1980}.

Last uncomputed quantity is the Coulomb logarithm. Because of quantum nature of
our scattering (see~\cite{ll10})
\begin{equation}
\frac{|ee'|}{\hbar v}\approx \frac{Ze^2}{\hbar c}=Z\alpha \ll 1,
\end{equation}
then the minimal momentum transferred is $q_{\rm min}=\hbar/l$, where $l$
is the typical
length of interaction. Then~(\ref{eq:CoulLog}) can be rewritten as
\begin{equation}
L_{ei}=\ln\left(\frac{2p_e l}{\hbar}\right)-\frac{v_F^2}{2c^2}.
\end{equation}
We should define $l$ -- maximum length with interaction. Recalling
reasoning from the previous paragraph we could put $l=a_i$, so in this case
the Coulomb logarithm:
\begin{equation}
L_{ei}=\ln\left((18\pi)^{1/3} Z^{1/3}\right)-\frac{v_F^2}{2c^2}
=\ln\left(3.838 Z^{1/3}\right)-\frac{v_F^2}{2c^2} .
\end{equation}
In hydrodynamics there are general expressions for mass and heat
fluxes (see, e.g.~\cite{ll6}):
\begin{eqnarray}
  {\bf i}&=&-\alpha\nabla\mu-\beta\nabla T,\nonumber\\
  {\bf q}&=&-\delta\nabla\mu-\gamma\nabla T.
  \label{sys:i_q}
\end{eqnarray}
Here we use a definition of ${\bf q}$ with subtraction of $\mu {\bf i}$.
Let us restore terms with $\nabla\mu$ in kinetic equation:
\begin{equation}
  -\frac{\partial f_0}{\partial\epsilon}\frac{\epsilon-\mu}{T}({\bf
    v}\nabla)T+\frac{\partial f_0}{\partial\epsilon}({\bf
    v}\nabla)\mu=
  \int w({\bf p},{\bf p'})(\phi-\phi')\frac{\partial
    f_0}{\partial\epsilon}d{\bf p'}.
  \label{eq:BolEq}
\end{equation}
Like above we look for solution in the form
\begin{equation}
  \phi=A(\epsilon-\mu)({\bf p}\nabla)T+B({\bf p}\nabla)\mu.
\end{equation}
And using the same techniques, we obtain
\begin{eqnarray}
  \phi&=&A(\epsilon-\mu)({\bf p}\nabla)T+AT({\bf
    p}\nabla)\mu,\nonumber\\
  A&=&-\frac{1}{4\pi}\frac{(\mu^2-m_e^2) v_F}{Z^2e^4n_i\mu \Lambda_{ei}T} .
\end{eqnarray}
Based on definitions
\begin{eqnarray}
  {\bf i}&=&-\int {\bf v}\frac{\partial f_0}{\partial\epsilon}\phi
  d{\bf p'},\nonumber\\
  {\bf q}&=&-\int (\epsilon-\mu){\bf v}\frac{\partial f_0}{\partial\epsilon}\phi
  d{\bf p'} ,
\end{eqnarray}
we obtain Onsager's result $\delta=T\beta$ and in our case (of
degenerate electrons)
$\alpha=-AT\frac{v_F^2}{3}\mu$; $\beta=0$; $\delta=0$; $\gamma=\kappa$.

At the end we can give some limits of applicability for our expression for
$\kappa$. As we can recall, we have used
\begin{equation}
\int\limits_0^\infty
\psi(\epsilon)\frac{\partial f_0}{\partial\epsilon}d\epsilon=
-\psi(\mu)
-\frac{\pi^2}{6}T^2\psi''(\mu)-\frac{7\pi^4}{360}T^4
\psi''''(\mu)+\ldots
\end{equation}
in our calculations. We have omitted the 3rd term and this is correct when
\begin{equation}
  T<\sqrt{\frac{60}{7\pi^2}\frac{\psi''(\mu)}{\psi''''(\mu)}}=0.16\mu,
\end{equation}
where the last equality arises from our previous calculations. This is
the validity criterion for our expression (\ref{eq:KappaRasm}).

\section{Conclusions}

This work presents methodical notes on Coulomb corrections in a
non-ideal plasma.
A quantity $a_i$ that enters the corrections in
different works is defined in a different manner, what can lead
sometimes to inaccurate results. We emphasize that a correct definition of
$a_i$ is a radius of electron cloud with charge $(-Z_i)$:
\begin{equation}
  a_i\equiv \left(\frac{3}{4\pi}\frac{Z_i}{n_e}\right)^{1/3}.
  \label{eq:final_a_i}
\end{equation}
This quantity is used for definition of free
energy~(\ref{eq:DeltaFC}). When the relation (\ref{eq:final_a_i}) is used the physical sense of correction to $F$
is clear: they arise due to interaction of ions with electron cloud,
and not ion-ion interactions.

An example of quantity that depends significantly on such corrections
is classical screening of nuclear reactions.
Sometimes chemical potentials are used for definition of $H_{12}(0)$.
We argue that a chemical potential does not reflect the essence of the matter:
the interaction is fully electromagnetic in a screening process,
and elements with different $Z$ have qualitatively the same nature relative to these interactions, while
no particles is taken from the system in nuclear reactions.
A correct formula should be written in terms of free energy and should read
\begin{equation}
  H_{12}(0)=-\left[F(N-2,1)-2F(N,0)\right]/T ,
\end{equation}
where $F(N-2,1)$ is a free energy for $N-2$ ions with charge $Z$ and one ion with charge $2Z$, while $F(N,0)$
is defined respectively for ions before reaction \cite{Jancovici1977}.

Finally, we give a straightforward derivation for a thermo-conductivity coefficient of
degenerate relativistic electron gas.

Authors would like to thank P.V. Sasorov and A.Yu. Potekhin for
helpful discussions,
and anonymous referees for valuable comments on the literature.
The work is supported partly by grants NSh-2977.2008.2 and
NSh-3884.2008.2, ``Dynasty'' foundation, and partly supported by Federal Programm "Scientific and
pedagogical specialists of innovation Russia.", contract number
02.740.11.0250.
SB is supported by MPA and Univ.~Tokyo guest programs
(through World Premier International Research Center Initiative, WPI, MEXT, Japan)
and partly by the Russian Foundation for Basic Research grant RFBR 07-02-00830-a.

\bibliographystyle{apsrev}
\bibliography{glazyrin_coulcond}

\begin{thebibliography}{26}
\expandafter\ifx\csname natexlab\endcsname\relax\def\natexlab#1{#1}\fi
\expandafter\ifx\csname bibnamefont\endcsname\relax
  \def\bibnamefont#1{#1}\fi
\expandafter\ifx\csname bibfnamefont\endcsname\relax
  \def\bibfnamefont#1{#1}\fi
\expandafter\ifx\csname citenamefont\endcsname\relax
  \def\citenamefont#1{#1}\fi
\expandafter\ifx\csname url\endcsname\relax
  \def\url#1{\texttt{#1}}\fi
\expandafter\ifx\csname urlprefix\endcsname\relax\def\urlprefix{URL }\fi
\providecommand{\bibinfo}[2]{#2}
\providecommand{\eprint}[2][]{\url{#2}}

\bibitem[{\citenamefont{Flowers and Itoh}(1976)}]{FlowersItoh1976}
\bibinfo{author}{\bibfnamefont{E.}~\bibnamefont{Flowers}} \bibnamefont{and}
  \bibinfo{author}{\bibfnamefont{N.}~\bibnamefont{Itoh}},
  \bibinfo{journal}{Astrophys. J.} \textbf{\bibinfo{volume}{206}},
  \bibinfo{pages}{218} (\bibinfo{year}{1976}).

\bibitem[{\citenamefont{Yakovlev and Urpin}(1980)}]{YakovlevUrpin1980}
\bibinfo{author}{\bibfnamefont{D.~G.} \bibnamefont{Yakovlev}} \bibnamefont{and}
  \bibinfo{author}{\bibfnamefont{V.~A.} \bibnamefont{Urpin}},
  \bibinfo{journal}{Soviet Astronomy} \textbf{\bibinfo{volume}{24}},
  \bibinfo{pages}{303} (\bibinfo{year}{1980}).

\bibitem[{\citenamefont{{Yakovlev}}(1987)}]{Yak1987}
\bibinfo{author}{\bibfnamefont{D.~G.} \bibnamefont{{Yakovlev}}},
  \bibinfo{journal}{Soviet Astronomy} \textbf{\bibinfo{volume}{31}},
  \bibinfo{pages}{347} (\bibinfo{year}{1987}).

\bibitem[{\citenamefont{{Cassisi} et~al.}(2007)\citenamefont{{Cassisi},
  {Potekhin}, {Pietrinferni}, {Catelan}, and {Salaris}}}]{CassisiEtAl2007}
\bibinfo{author}{\bibfnamefont{S.}~\bibnamefont{{Cassisi}}},
  \bibinfo{author}{\bibfnamefont{A.~Y.} \bibnamefont{{Potekhin}}},
  \bibinfo{author}{\bibfnamefont{A.}~\bibnamefont{{Pietrinferni}}},
  \bibinfo{author}{\bibfnamefont{M.}~\bibnamefont{{Catelan}}},
  \bibnamefont{and}
  \bibinfo{author}{\bibfnamefont{M.}~\bibnamefont{{Salaris}}},
  \bibinfo{journal}{\apj} \textbf{\bibinfo{volume}{661}}, \bibinfo{pages}{1094}
  (\bibinfo{year}{2007}), \eprint{arXiv:astro-ph/0703011}.

\bibitem[{\citenamefont{{Ichimaru}}(1993)}]{Ichimaru1993}
\bibinfo{author}{\bibfnamefont{S.}~\bibnamefont{{Ichimaru}}},
  \bibinfo{journal}{Reviews of Modern Physics} \textbf{\bibinfo{volume}{65}},
  \bibinfo{pages}{255} (\bibinfo{year}{1993}).

\bibitem[{\citenamefont{{Kraeft} et~al.}(1986)\citenamefont{{Kraeft}, {Kremp},
  {Ebeling}, and {R\"opke}}}]{Kraeft1986}
\bibinfo{author}{\bibfnamefont{W.-D.} \bibnamefont{{Kraeft}}},
  \bibinfo{author}{\bibfnamefont{D.}~\bibnamefont{{Kremp}}},
  \bibinfo{author}{\bibfnamefont{W.}~\bibnamefont{{Ebeling}}},
  \bibnamefont{and}
  \bibinfo{author}{\bibfnamefont{G.}~\bibnamefont{{R\"opke}}},
  \emph{\bibinfo{title}{{Quantum Statistics of Charged Particle Systems}}}
  (\bibinfo{publisher}{Akademle-Verlag Berlin}, \bibinfo{year}{1986}).

\bibitem[{\citenamefont{{Pain} et~al.}(2006)\citenamefont{{Pain}, {Dejonghe},
  and {Blenski}}}]{PDB06}
\bibinfo{author}{\bibfnamefont{J.~C.} \bibnamefont{{Pain}}},
  \bibinfo{author}{\bibfnamefont{G.}~\bibnamefont{{Dejonghe}}},
  \bibnamefont{and}
  \bibinfo{author}{\bibfnamefont{T.}~\bibnamefont{{Blenski}}},
  \bibinfo{journal}{Journal of Quantitative Spectroscopy and Radiative
  Transfer} \textbf{\bibinfo{volume}{99}}, \bibinfo{pages}{451}
  (\bibinfo{year}{2006}).

\bibitem[{\citenamefont{{Dewitt} and {Slattery}}(1999)}]{DewittSlattery1999}
\bibinfo{author}{\bibfnamefont{H.}~\bibnamefont{{Dewitt}}} \bibnamefont{and}
  \bibinfo{author}{\bibfnamefont{W.}~\bibnamefont{{Slattery}}},
  \bibinfo{journal}{Contributions to Plasma Physics}
  \textbf{\bibinfo{volume}{39}}, \bibinfo{pages}{97} (\bibinfo{year}{1999}).

\bibitem[{\citenamefont{{Mazurek} et~al.}(1979)\citenamefont{{Mazurek},
  {Lattimer}, and {Brown}}}]{maz79}
\bibinfo{author}{\bibfnamefont{T.~J.} \bibnamefont{{Mazurek}}},
  \bibinfo{author}{\bibfnamefont{J.~M.} \bibnamefont{{Lattimer}}},
  \bibnamefont{and} \bibinfo{author}{\bibfnamefont{G.~E.}
  \bibnamefont{{Brown}}}, \bibinfo{journal}{\apj}
  \textbf{\bibinfo{volume}{229}}, \bibinfo{pages}{713} (\bibinfo{year}{1979}).

\bibitem[{\citenamefont{Yakovlev and Shalybkov}(1989)}]{yak89}
\bibinfo{author}{\bibfnamefont{D.~G.} \bibnamefont{Yakovlev}} \bibnamefont{and}
  \bibinfo{author}{\bibfnamefont{D.~A.} \bibnamefont{Shalybkov}},
  \bibinfo{journal}{Astrophysics and Space Physics Reviews, Soviet Sci.\ Rev.}
  \textbf{\bibinfo{volume}{E7}}, \bibinfo{pages}{311} (\bibinfo{year}{1989}).

\bibitem[{\citenamefont{{Landau} and {Lifshitz}}(2002{\natexlab{a}})}]{ll10}
\bibinfo{author}{\bibfnamefont{L.~D.} \bibnamefont{{Landau}}} \bibnamefont{and}
  \bibinfo{author}{\bibfnamefont{E.~M.} \bibnamefont{{Lifshitz}}},
  \emph{\bibinfo{title}{{Physical kinetics}}} (\bibinfo{publisher}{Moskva
  Fizmatlit}, \bibinfo{year}{2002}{\natexlab{a}}).

\bibitem[{\citenamefont{Haensel et~al.}(2007)\citenamefont{Haensel, Potekhin,
  and Yakovlev}}]{HPYa07}
\bibinfo{author}{\bibfnamefont{P.}~\bibnamefont{Haensel}},
  \bibinfo{author}{\bibfnamefont{A.~Y.} \bibnamefont{Potekhin}},
  \bibnamefont{and} \bibinfo{author}{\bibfnamefont{D.~G.}
  \bibnamefont{Yakovlev}}, \emph{\bibinfo{title}{Neutron stars; 1 Equation of
  state and structure}}, Astrophysics and Space Science Library
  (\bibinfo{publisher}{Springer-Verlag}, \bibinfo{address}{New York},
  \bibinfo{year}{2007}).

\bibitem[{\citenamefont{{Brush} et~al.}(1966)\citenamefont{{Brush}, {Sahlin},
  and {Teller}}}]{BST1966}
\bibinfo{author}{\bibfnamefont{S.~G.} \bibnamefont{{Brush}}},
  \bibinfo{author}{\bibfnamefont{H.~L.} \bibnamefont{{Sahlin}}},
  \bibnamefont{and} \bibinfo{author}{\bibfnamefont{E.}~\bibnamefont{{Teller}}},
  \bibinfo{journal}{J. Chem. Phys.} \textbf{\bibinfo{volume}{45}},
  \bibinfo{pages}{2102} (\bibinfo{year}{1966}).

\bibitem[{\citenamefont{{Dewitt} et~al.}(1973)\citenamefont{{Dewitt},
  {Graboske}, and {Cooper}}}]{DewittGraboskeCooper1973}
\bibinfo{author}{\bibfnamefont{H.~E.} \bibnamefont{{Dewitt}}},
  \bibinfo{author}{\bibfnamefont{H.~C.} \bibnamefont{{Graboske}}},
  \bibnamefont{and} \bibinfo{author}{\bibfnamefont{M.~S.}
  \bibnamefont{{Cooper}}}, \bibinfo{journal}{\apj}
  \textbf{\bibinfo{volume}{181}}, \bibinfo{pages}{439} (\bibinfo{year}{1973}).

\bibitem[{\citenamefont{{Brown} et~al.}(2006)\citenamefont{{Brown}, {Dooling},
  and {Preston}}}]{Brown06}
\bibinfo{author}{\bibfnamefont{L.~S.} \bibnamefont{{Brown}}},
  \bibinfo{author}{\bibfnamefont{D.~C.} \bibnamefont{{Dooling}}},
  \bibnamefont{and} \bibinfo{author}{\bibfnamefont{D.~L.}
  \bibnamefont{{Preston}}}, \bibinfo{journal}{\pre}
  \textbf{\bibinfo{volume}{73}} (\bibinfo{year}{2006}),
  \eprint{arXiv:physics/0603250}.

\bibitem[{\citenamefont{{Brydges} and {Martin}}(1999)}]{Brydges99}
\bibinfo{author}{\bibfnamefont{D.~C.} \bibnamefont{{Brydges}}}
  \bibnamefont{and} \bibinfo{author}{\bibfnamefont{P.~A.}
  \bibnamefont{{Martin}}}, \bibinfo{journal}{Journal of Statistical Physics}
  \textbf{\bibinfo{volume}{96}}, \bibinfo{pages}{1163} (\bibinfo{year}{1999}),
  \eprint{arXiv:cond-mat/9904122}.

\bibitem[{\citenamefont{{Blinnikov} et~al.}(2009)\citenamefont{{Blinnikov},
  {Panov}, {Rudzsky}, and {Sumiyoshi}}}]{BPRS2009}
\bibinfo{author}{\bibfnamefont{S.~I.} \bibnamefont{{Blinnikov}}},
  \bibinfo{author}{\bibfnamefont{I.~V.} \bibnamefont{{Panov}}},
  \bibinfo{author}{\bibfnamefont{M.~A.} \bibnamefont{{Rudzsky}}},
  \bibnamefont{and}
  \bibinfo{author}{\bibfnamefont{K.}~\bibnamefont{{Sumiyoshi}}},
  \bibinfo{journal}{ArXiv e-prints}  (\bibinfo{year}{2009}),
  \eprint{0904.3849}.

\bibitem[{\citenamefont{{Nadyozhin} and {Yudin}}(2005)}]{nad05}
\bibinfo{author}{\bibfnamefont{D.~K.} \bibnamefont{{Nadyozhin}}}
  \bibnamefont{and} \bibinfo{author}{\bibfnamefont{A.~V.}
  \bibnamefont{{Yudin}}}, \bibinfo{journal}{Astronomy Letters}
  \textbf{\bibinfo{volume}{31}}, \bibinfo{pages}{271} (\bibinfo{year}{2005}).

\bibitem[{\citenamefont{{Jancovici}}(1977)}]{Jancovici1977}
\bibinfo{author}{\bibfnamefont{B.}~\bibnamefont{{Jancovici}}},
  \bibinfo{journal}{Journal of Statistical Physics}
  \textbf{\bibinfo{volume}{17}}, \bibinfo{pages}{357} (\bibinfo{year}{1977}).

\bibitem[{\citenamefont{{Hillebrandt} and
  {Niemeyer}}(2000)}]{HillebrandtNiemeyer2000}
\bibinfo{author}{\bibfnamefont{W.}~\bibnamefont{{Hillebrandt}}}
  \bibnamefont{and} \bibinfo{author}{\bibfnamefont{J.~C.}
  \bibnamefont{{Niemeyer}}}, \bibinfo{journal}{Annual Review of Astronomy and
  Astrophysics} \textbf{\bibinfo{volume}{38}}, \bibinfo{pages}{191}
  (\bibinfo{year}{2000}), \eprint{arXiv:astro-ph/0006305}.

\bibitem[{\citenamefont{{van Horn}}(1969)}]{vHorn1969}
\bibinfo{author}{\bibfnamefont{H.~M.} \bibnamefont{{van Horn}}},
  \bibinfo{journal}{Physics Letters A} \textbf{\bibinfo{volume}{28}},
  \bibinfo{pages}{706} (\bibinfo{year}{1969}).

\bibitem[{\citenamefont{{Slattery} et~al.}(1980)\citenamefont{{Slattery},
  {Doolen}, and {Dewitt}}}]{SlDooDew1980}
\bibinfo{author}{\bibfnamefont{W.~L.} \bibnamefont{{Slattery}}},
  \bibinfo{author}{\bibfnamefont{G.~D.} \bibnamefont{{Doolen}}},
  \bibnamefont{and} \bibinfo{author}{\bibfnamefont{H.~E.}
  \bibnamefont{{Dewitt}}}, \bibinfo{journal}{\pra}
  \textbf{\bibinfo{volume}{21}}, \bibinfo{pages}{2087} (\bibinfo{year}{1980}).

\bibitem[{\citenamefont{{Jones} and {Ceperley}}(1996)}]{JP1996}
\bibinfo{author}{\bibfnamefont{M.~D.} \bibnamefont{{Jones}}} \bibnamefont{and}
  \bibinfo{author}{\bibfnamefont{D.~M.} \bibnamefont{{Ceperley}}},
  \bibinfo{journal}{\prl} \textbf{\bibinfo{volume}{76}}, \bibinfo{pages}{4572}
  (\bibinfo{year}{1996}).

\bibitem[{\citenamefont{{Landau} and {Lifshitz}}(2002{\natexlab{b}})}]{ll4}
\bibinfo{author}{\bibfnamefont{L.~D.} \bibnamefont{{Landau}}} \bibnamefont{and}
  \bibinfo{author}{\bibfnamefont{E.~M.} \bibnamefont{{Lifshitz}}},
  \emph{\bibinfo{title}{{Quantum electrodynamics}}} (\bibinfo{publisher}{Moskva
  Fizmatlit}, \bibinfo{year}{2002}{\natexlab{b}}).

\bibitem[{\citenamefont{Ziman}(1960)}]{Ziman1960}
\bibinfo{author}{\bibfnamefont{J.~M.} \bibnamefont{Ziman}},
  \emph{\bibinfo{title}{Electrons and Phonons}} (\bibinfo{publisher}{Oxford
  University Press}, \bibinfo{year}{1960}).

\bibitem[{\citenamefont{{Landau} and {Lifshitz}}(1986)}]{ll6}
\bibinfo{author}{\bibfnamefont{L.~D.} \bibnamefont{{Landau}}} \bibnamefont{and}
  \bibinfo{author}{\bibfnamefont{E.~M.} \bibnamefont{{Lifshitz}}},
  \emph{\bibinfo{title}{{Hydrodynamics}}} (\bibinfo{publisher}{Moskva
  Fizmatlit}, \bibinfo{year}{1986}).

\end{thebibliography}

\end{document}